\documentclass[%
 aip,
 jmp,%
 amsmath,amssymb,
reprint,%
]{revtex4-1}

\usepackage{graphicx}
\usepackage{dcolumn}
\usepackage{bm}
\usepackage{color}

\begin{document}

\preprint{AIP/123-QED}

\title{Eigenvalues of normalized Laplacian matrices of fractal trees and dendrimers: Analytical results and applications}

\author{Alafate Julaiti}
\author{Bin Wu}
\author{Zhongzhi Zhang}
\email{zhangzz@fudan.edu.cn}
\homepage{http://www.researcherid.com/rid/G-5522-2011}

\affiliation {School of Computer Science, Fudan University,
Shanghai 200433, China}

\affiliation {Shanghai Key Lab of Intelligent Information
Processing, Fudan University, Shanghai 200433, China}

\date{\today}

\begin{abstract}
The eigenvalues of the normalized Laplacian matrix of a network plays an important role in its structural and dynamical aspects associated with the network. In this paper,
we study the spectra and their applications of normalized Laplacian matrices of a family of fractal trees and dendrimers modeled by Cayley trees, both of which are built in an iterative way. For the fractal trees, we apply the spectral decimation approach to determine analytically all the eigenvalues and their corresponding multiplicities, with the eigenvalues provided by a recursive relation governing the eigenvalues of networks at two successive generations. For Cayley trees, we show that all their eigenvalues can be obtained by computing the roots of several small-degree polynomials defined recursively. By using the relation between normalized Laplacian spectra and eigentime identity, we derive the explicit solution to the eigentime identity for random walks on the two treelike networks, the leading scalings of which follow quite different behaviors. In addition, we corroborate the obtained eigenvalues and their degeneracies through the link between them and the number of spanning trees.
\end{abstract}


\maketitle


\section{Introduction}

A central issue in the study of complex systems is to understand the topological structure and to further unveil how various structural properties affect the dynamical processes occurring on diverse systems~\cite{GuBl05,BoLaMoChHw06,DoGoMe08}. It has been established that numerous structural and dynamical properties of a networked system are encoded in eigenvalues and eigenvectors of its standard Laplacian matrix. Frequently cited examples include resistance distance~\cite{Wu04}, relaxation dynamic in the framework of generalized Gaussian structure~\cite{GuBl05,SoBl95,Sc98}, fluorescence depolarization by quasiresonant energy transfer~\cite{BlVoJuKo05JOL,BlVoJuKo05,LiuZh13}, continuous-time quantum walks~\cite{AhDaZa93,AgBlMu08,MuBl11}, and so on.
Therefore, it is of extreme importance to study the spectra of standard Laplacian matrices of complex systems. Thus far, the eigenvalues of standard Laplacian matrices for some particular graphs have been found exactly, including regular hypercubic lattices~\cite{GuBl05,DejoMe98}, dual Sierpinski gaskets~\cite{CoKa92,MaMaPe97,BlJu02}, Vicsek fractals~\cite{JaWuCo92,JaWu94}, Vicsek fractals replicated in the shape of dual Sierpinski gaskets~\cite{JuVoBe11}, dendrimers~\cite{CaCh97} and their dual graphs~\cite{GaBl07,Ga10}, as well as some small-world networks with a degree distribution of exponential form~\cite{ZhZhQiGu08,ZhQiZhLiGu09,GrGrTi12}. These works presented novel approaches and paradigms for the computation about spectra of standard Laplacian matrices.

In addition to the spectrum of standard Laplacian matrix, the eigenvalues and eigenvectors of normalized Laplacian matrix of a network also contain much important information about its architecture and dynamical processes. In the structural aspect, for example, the product of all nonzero eigenvalues of a connected network determines the number of spanning trees in the network~\cite{Ch97}; the nonzero eigenvalues and their orthonormalized eigenvectors can be used to express the resistor resistance between any pair of nodes~\cite{ChZh07}. With respect to dynamical processes, many interesting quantities of random walks are related to the eigenvalues and eigenvectors of normalized Laplacian matrix, including mean first-passage time~\cite{Lo96,AlFi99,ZhShCh13}, mixing time~\cite{Lo96,AlFi99}, and Kemeny constant~\cite{KeSn76} or eigentime identity~\cite{AlFi99} that can be used as a measure of efficiency of navigation on the network~\cite{LeLo02}. Particularly, eigenvalues and eigenvectors of normalized Laplacian matrix are relevant in light harvesting~\cite{BaKlKo97}, energy or exciton transport~\cite{BlZu81,Kn68}, chemical kinetics~\cite{Ki58,MoSh57} and many other problems in chemical physics~\cite{We67}.

In view of the relevance, it is equally important to compute and analyze the spectra of normalized Laplacian matrices. However, since normalized Laplacian matrix and standard Laplacian matrix of a network behave quite differently~\cite{ChDaHaLiPaSt04}, the spectra of one matrix can be determined does not mean that the spectra of the other matrix can also be evaluated. For instance, the eigenvalues of Laplacian matrix of Vicsek fractals can be determined analytically~\cite{BlJuKoFe03,BlFeJuKo04,ZhWjZhZhGuWa10}, but it is difficult (maybe impossible) to derive their spectra of normalized Laplacian matrix in a similar way. Thus, the spectra of standard Laplacian matrix and normalized Laplacian matrix must be treated individually. Relative to standard Laplacian matrices, the spectra of normalized Laplacian matrices have received little attention~\cite{ChLuVu03,HwYuLeKa10,KiKa07,WuZh12}.

In this paper, we present a theoretical study of the normalized Laplacian matrices for a family of proposed fractal trees and Cayley trees~\cite{CaCh97,ChCa99} as a classic model of dendrimers. Both networks are constructed in an iterative way, which makes it possible to analytically study the spectra of their normalized Laplacian matrices. For the fractal trees, by making use of the spectrum decimation approach~\cite{DoAlBeKa83,Ra84}, we deduce a recursion relation for the eigenvalues at every two successive iterations, and derive the multiplicity of each eigenvalue. For Cayley trees, the problem of computing eigenvalue spectra is reduced to determining the roots of some small-degree polynomials that are defined recursively.

As an application, we further derive closed-form expressions of eigentime identity for random walks in both networks being studied, by using the obtained recursive relations for eigenvalues and polynomials. The eigentime identity can be looked upon as the trapping efficiency for a kind of particular trapping process~\cite{AlFi99}. The results show that the dominating terms of the Kemeny constants for the two networks display strongly different behaviors, indicating the effects of topologies on the navigation efficiency. Furthermore, we test the validity of the eigenvalue computations using the connection between the number of spanning trees and the product of nonzero eigenvlaues of normalized Laplacian matrix of a network.

\begin{figure}
\begin{center}
\includegraphics[width=0.8\linewidth]{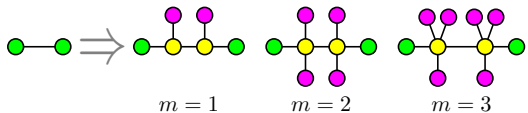}
\caption{(Color online) Construction method of the
fractal trees. The next generation of the fractal trees can be obtained through replacing each edge of the present generation by the clusters on the right-hand side of the arrow.}
\label{cons}
\end{center}
\end{figure}

\section{Network constructions and properties}\label{Constr}

In this section, we introduce the constructions and some relevant properties of a family of fractal trees and Cayley trees. Both networks are constructed in an iterative manner. The special
constructions allow us to treat analytically their properties and dynamical processes taking place on them.

\subsection{Fractal trees\label{SpecF}}

\begin{figure}
\begin{center}
\includegraphics[width=0.95\linewidth,trim=90 0 0 0]{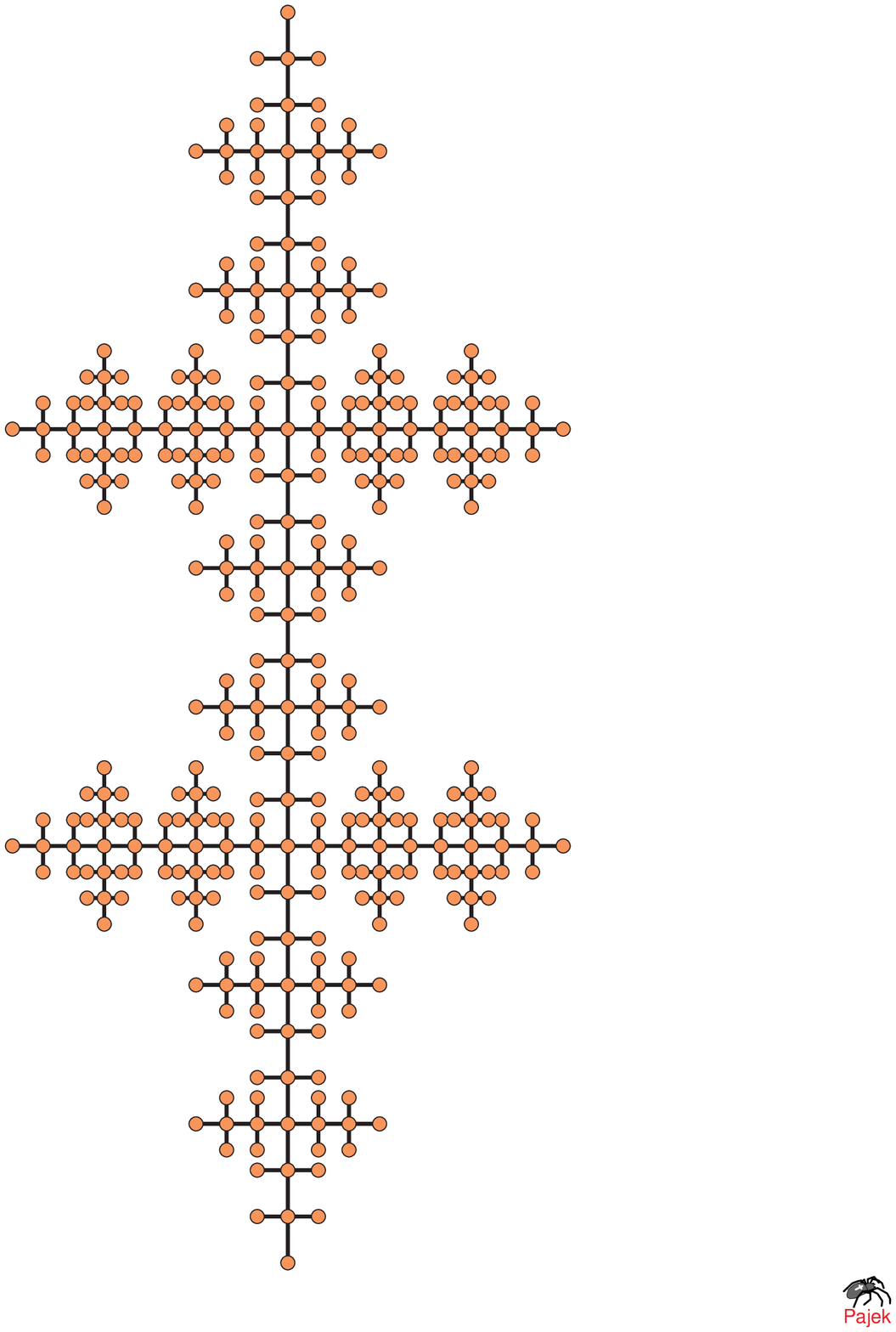}
\end{center}
\caption[kurzform]{(Color online) Illustration of a
special network $F_3$ for the case of $m=2$.} \label{Fractal}
\end{figure}

We first introduce the construction and properties of a family of fractal trees. Let $F_n$ ($n\geq 0$) denote the proposed fractal trees after $n$ iterations. For $n=0$, $F_0$ is an edge connecting two nodes. For $n \geq 1$, $F_n$ is obtained from $F_{n-1}$ by performing two operations on each existing edge in $F_{n-1}$ as shown in Fig.~\ref{cons}. The first operation is to replace the edge by a path of 3 links long, with the two endpoints of the path being the same endpoints of the original edge. The second operation is to create $m$ (a positive integer) new nodes for each of two middle nodes in the path, and attach them to the middle node. Figure~\ref{Fractal} illustrates a network $F_3$ corresponding to a particular case of $m=2$.

By construction, at each generation $n_i$ ($n_i \geq 1$), the number of newly introduced nodes is $2(m+1)$ times the number of edges at generation $n_i - 1$. Let $N_n$ and $E_n$ denote, respectively, the number of nodes and the number of edges in $F_n$. Then, $N_n$ and $E_n$ obey the following relations:
\begin{equation}\label{Nn_recur}
N_n=N_{n-1} + 2(m+1)E_{n-1}
\end{equation}
and
\begin{equation}\label{En_recur}
E_n=N_{n} -1.
\end{equation}
Considering $N_0 = 2$ and $E_0 = 1$, Equations~(\ref{Nn_recur}) and~(\ref{En_recur}) are solved to obtain:
\begin{equation}\label{Nn}
N_n=(2m+3)^{n}+1
\end{equation}
and
\begin{equation}\label{En}
E_n=(2m+3)^{n}\,.
\end{equation}

Equation~(\ref{Nn}) shows that after the evolution of one generation, the number of nodes increases by a factor $f_N=2m+3$. In addition, it it easy to check that after each iteration the diameter increases by a factor of  $f_D=3$. Thus, the fractal dimension of the trees is $f_B=\ln f_N/ \ln f_D=  \ln (2m+3)/ \ln 3$. Furthermore, the fractal trees are ``large-world'' with their diameter growing in a power of the network size as $(N_n)^{\ln 3/\ln (2m+3)}$.

\subsection{Cayley trees}\label{CayleyT}

\begin{figure}
\begin{center}
\includegraphics[width=0.85\linewidth,trim=0 0 0 0]{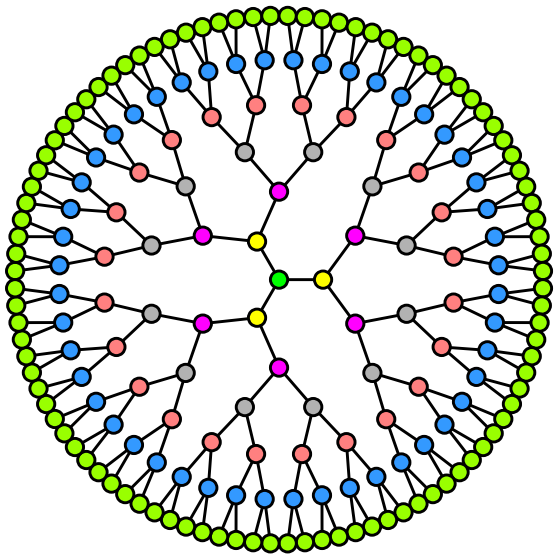}
\caption{(Color online) The Cayley tree $C_{3,6}$.}\label{CTree}
\end{center}
\end{figure}

Let $C_{b,n}$ ($b \geq 3$, $n \geq 0$) represent the Cayley trees after $n$ iterations (generations), which can be built in the following iterative way~\cite{CaCh97,ChCa99}. Initially ($n=0$), $C_{b,0}$ consists of only a central node. To form $C_{b,1}$, we create $b$ nodes and attach them to the central node. For any $n>1$, $C_{b,n}$ is obtained
from $C_{b,n-1}$ by performing the following operation. For each boundary node of $C_{b,n-1}$, $b-1$ nodes are generated and attached to the boundary node. Figure~\ref{CTree} illustrates a special Cayley tree, $C_{3,6}$.

Let $N_{i}(n)$ denote the number of nodes in
$C_{b,n}$, which are given birth to at iteration $i$. It is easy to check that
\begin{equation}\label{cay1}
N_{i}(n)=\begin{cases}
1, &i=0,\\
b(b-1)^{i-1}, &i>0.\\
\end{cases}
\end{equation}
Thus, the network size of $C_{b,n}$ is
\begin{equation}\label{cay1a}
N_n=\sum_{i=0}^{n}N_{i}(n)=\frac{b(b-1)^n-2}{b-2}\,.
\end{equation}
Different from the fractal trees introduced above, Cayley trees are nonfractal, irrespective of their self-similar architectures.
That is, the fractal dimension of Cayley trees is infinite.

After introducing the constructions and features of the fractal trees and Cayley trees, we next  study the eigenvalue spectrum of their normalized Laplacian matrices.

\section{Normalized Laplacian spectrum of the fractal trees}

We now address the spectra of normalized Laplacian matrix of the fractal trees, for which we will give a recursive solution to eigenvalues and determine the degeneracy of each eigenvalue. Moreover, we will use the obtained eigenvalues to compute the eigentime identity of random walks, as well as the number of spanning trees.

\subsection{Eigenvalue spectrum}

As is known to us all, the structure of $F_{n}$ is encoded in its adjacency matrix $A_n$, the entry $A_n(i,j)$ of which is $1$ (or $0$) if nodes $i$ and $j$ are (not) adjacent in $F_{n}$. Then, the normalized Laplacian matrix of $F_{n}$, denoted by $P_n$, is defined as $P_n=I_n-D_n^{-\frac{1}{2}} A_n D_n^{-\frac{1}{2}}$, where $I_n$ is the identity matrix of the same order as $A_n$ and $D_n$ is the diagonal degree matrix of $F_{n}$ with its $i$th diagonal entry being the degree, $d_i(n)$, of node $i$ in $F_{n}$. Notice that $D_n^{-\frac{1}{2}} A_n D_n^{-\frac{1}{2}}$ is similar to the Markov matrix $M_n=D_n^{-1}A_n$ of $F_{n}$, which can be seen from the equivalent relation $D_n^{-\frac{1}{2}} A_n D_n^{-\frac{1}{2}}=D_n ^{\frac{1}{2}}M_n D_n^{-\frac{1}{2}}$. Hence, both $D_n^{-\frac{1}{2}} A_n D_n^{-\frac{1}{2}}$ and $M_n$ have the same set of eigenvalues, although the former is symmetric, while the latter is often asymmetric. Note that if $\sigma_{i}(n)$ is an eigenvalue of $M_n$, then $1-\sigma_{i}(n)$ is an eigenvalue of the normalized Laplacian matrix of $P_n$. This one-to-one corresponding relation shows that if one can compute the eigenvalues of one matrix, the eigenvalues of another matrix can be easily found.

Let's examine the eigenvalue spectrum of $P_n$. It is easy to verify that the $(i,j)$ entry of $P_n$ is $P_n(i,j)=I_n(i,j)-\frac{A_n(i,j)}{\sqrt{d_i(n)}\sqrt{d_j(n)}}$, where $I_n(i,j)$ is the $(i,j)$ entry of $I_n$. Next we use the decimation technique~\cite{DoAlBeKa83,Ra84} to find the eigenvalues of $P_n$. The decimation method is general and has been applied to calculate the spectra of standard Laplacian matrices of Vicsek fractals~\cite{BlJuKoFe03,BlFeJuKo04,ZhWjZhZhGuWa10} and their extensions~\cite{JuVoBe11}, as well as dual Sierpinski gaskets~\cite{CoKa92,MaMaPe97,BlJu02}.

In order to find the recursive solutions to the normalized Laplacian spectra of the fractal trees, we now address the eigenvalue problem for matrix $P_{n+1}$.
Let $\alpha$ denote the set of original nodes belonging to $F_{n}$, and $\beta$ the set of nodes created at $(n+1)$th iteration. Then, $P_{n+1}$ can be written in the following block form
\begin{equation}\label{T2}
P_{n+1}=\left[\begin{array}{cccc}
P_{\alpha,\alpha} & P_{\alpha, \beta} \\
P_{\beta,\alpha} & P_{\beta, \beta}
\end{array}
\right]
=\left[\begin{array}{cccc}
I_{n} & P_{\alpha, \beta} \\
P_{\beta,\alpha} & P_{\beta, \beta}
\end{array}
\right]\,,
\end{equation}
where the block matrix $P_{\alpha,\alpha}$ describes transition probability between any pair of original nodes in $\alpha$, which is exactly equal to the identity matrix $I_{n}$ with order $N_n \times N_n$, $P_{\beta, \beta}$ explains the transition probability between any two nodes in $\beta$, which is block diagonal as will be shown detailedly in the following, and $P_{\alpha,\beta}$ (resp. $P_{\beta,\alpha}$) depicts the transition probability from any node in $\alpha$ (resp. $\beta$) to another node in $\beta$ (resp. $\alpha$).

Suppose $\lambda_{i}(n+1)$ is an eigenvalue of $P_{n+1}$, and $u=(u_{\alpha},u_{\beta})^\top$ is its corresponding eigenvector, where the superscript $\top$ stands for transpose and $u_{\alpha}$ and $u_{\beta}$ correspond to nodes in $\alpha$ and $\beta$, respectively.
Then, one can write the eigenvalue equation for matrix $P_{n+1}$ in the following block form:
\begin{equation}\label{T1}
\left[\begin{array}{cccc}
I_{n} & P_{\alpha, \beta} \\
P_{\beta,\alpha} & P_{\beta, \beta}
\end{array}
\right]
\left[\begin{array}{cccc}
 u_{\alpha} \\
 u_{\beta}
\end{array}
\right]={\lambda}_{i}(n+1) \left[\begin{array}{cccc}
 u_{\alpha} \\
 u_{\beta}
\end{array}
\right].
\end{equation}
Equation~(\ref{T1}) can be recast into two equations:
\begin{equation}\label{T3}
I_{n}u_{\alpha}+P_{\alpha, \beta}u_{\beta}=\lambda_{i}(n+1)u_{\alpha},
\end{equation}
\begin{equation}\label{T4}
P_{\beta,\alpha}u_{\alpha}+P_{\beta,\beta}u_{\beta}=\lambda_{i}(n+1)u_{\beta}.
\end{equation}
Equation~(\ref{T4}) implies
\begin{equation}\label{T5}
u_{\beta}=[\lambda_{i}(n+1)-P_{\beta,\beta}]^{-1}P_{\beta,\alpha}u_{\alpha}\,,
\end{equation}
provided that matrix $\lambda_{i}(n+1)-P_{\beta,\beta}$ is reversible. Inserting Eq.~(\ref{T5})
into Eq.~(\ref{T3}) yields
\begin{equation}\label{T6}
\{I_{n}+P_{\alpha,\beta}\left[\lambda_i(n+1)-P_{\beta,\beta}\right]^{-1}P_{\beta,\alpha}\}u_\alpha = \lambda_i(n+1)u_\alpha\,.
\end{equation}

Let $T_n =[\lambda_i(n + 1)-P_{\beta,\beta}]^{-1}$, the $(i,j)$th element of which is represented by $T_n(i,j)$, and let $Q_n =I_{n}+P_{\alpha,\beta}[\lambda_i(n+1)-P_{\beta,\beta}]^{-1}P_{\beta,\alpha}=I_{n}+P_{\alpha,\beta}T_{n}P_{\beta,\alpha}$. By construction, each edge in $F_{n}$ will generate $2(m+1)$ new nodes belonging to $F_{n+1}$, among which two nodes (denoted as $k$ and $l$) have a degree of $m+2$, the other $2m$ nodes have a single degree. In the Appendix, we prove that
\begin{equation}\label{T7}
Q_n=-\frac{T_n(l,k)}{m+2} P_n+\left(1+\frac{T_n(l,l)+T_n(l,k)}{m+2}\right)I_n\,,
\end{equation}
which related $Q_n$ to $P_{n}$, and thus enables one to represent the eigenvalues of matrix $P_{n+1}$ in terms of those of $P_{n}$.

Substituting Eq.~(\ref{T7}) into Eq.~(\ref{T6}) yields
\begin{small}
\begin{equation}\label{T10}
\left[\left(1+\frac{T_n(l,l)+T_n(l,k)}{m+2}\right) I_n -\frac{T_n(l,k)}{m+2} P_n \right] u_\alpha = \lambda_i(n+1) u_\alpha \,.
\end{equation}
\end{small}
Define $x = 1+\frac{T_n(l,l)+T_n(l,k)}{m+2}$ and $y = -\frac{T_n(l,k)}{m+2}$.
Equation~(\ref{T10}) can be rewritten as
\begin{equation}\label{T12}
P_n u_{\alpha}=\frac{\lambda_i(n+1)-x}{y}u_{\alpha}\,.
\end{equation}
Thus, if $\lambda_{i}(n)$ is an eigenvalue of $P_{n}$ with associated eigenvector $u_{\alpha}$, then
\begin{equation}\label{T13}
\lambda_{i}(n)=\frac{\lambda_i{(n+1)}-x}{y}\,.
\end{equation}
Plugging the results for $T_n(l,l)$ and $T_n(l,k)$ obtained in the Appendix into Eq.~(\ref{T13}), we obtain
\begin{eqnarray}\label{T16}
&(m+2)^2 [\lambda_i(n+1)] ^4-4 (m+2)^2 [\lambda_i(n+1)] ^3\nonumber\\
&  +\left(4 m^2+19 m+21\right) [\lambda_i(n+1)] ^2\nonumber\\
& -[6 m+9+\lambda_i(n)] [\lambda_i(n+1)] +\lambda_i(n)=0\,.
\end{eqnarray}
Equation~(\ref{T16}) relates $\lambda_{i}(n+1)$ to $\lambda_{i}(n)$, with each $\lambda_{i}(n)$ producing four eigenvalues of $P_{n+1}$.

\subsection{Multiplicities of eigenvalues}

In order determine the multiplicity of each eigenvalue, we first calculate numerically the eigenvalues for some networks of small sizes. The eigenvalues of $F_0$ are $0$ and $2$; the eigenvalues of $F_1$ are $0$, $2$, $\frac{1}{m+2}$, $\frac{2m+3}{m+2}$ and $1$: eigenvalue $1$ has a degeneracy of $2m$, while the other four eigenvalues have a single degeneracy. For $n\geq 2$,  the eigenvalue spectrum of $F_n$ exhibits the following properties: (i) Eigenvalues $0$, $2$, $\frac{1}{m+2}$ and $\frac{2m+3}{m+2}$ are present at any iteration, each having a single degeneracy. (ii) All eigenvalues present at a given generation $n_i$ will always exist at its
succeeding generation $n_i+1$, and all new eigenvalues at generation $n_{i}+1$ are just those generated via Eq.~(\ref{T16}) by substituting $\lambda_{i}(n_i)$ ($\lambda_{i}(n_i) \neq 0,2$) for $\lambda_{i}(n)$, where $\lambda_{i}(n_i)$ are those eigenvalues that are newly added to generation $n_{i}$; moreover each new eigenvalue at generation $n_{i}+1$ keeps the degeneracy of its father.

Using property (ii) of eigenvalues, we now determine the degeneracy of eigenvalue $1$ and the multiplicities of its offsprings. Let $D^{\rm mul}_n(\lambda)$ represent the multiplicity of eigenvalue $\lambda$ of matrix $P_n$. And let $r(M)$ denote the rank of matrix $M$. Then, the multiplicity of eigenvalue $1$ of $P_{n}$ is
\begin{equation}\label{N0}
D^{\rm mul}_{n}(\lambda=1)= N_{n}-r(P_{n}-I_n)\,.
\end{equation}
Thus, to determine $D^{\rm mul}_{n}(\lambda=1)$, we can alternatively compute $r(P_{n}-I_n)$. By using some elementary matrix operations, we can eliminate all nonzero elements at $P_{\alpha,\beta}$, leading to $r(P_{n}-I_n)=r(P_{\beta,\beta}-I_{\beta,\beta})$, where $I_{\beta,\beta}$ is the identity matrix with the same order as $P_{\beta,\beta}$. By construction, it is easy to see that $r(P_{\beta,\beta}-I_{\beta,\beta})$ is block diagonal with the rank of each of $(2m+3)^{n-1}$ blocks being $4$, then $r(P_{\beta,\beta}-I_{\beta,\beta})=4(2m+3)^{n-1}$ and
\begin{equation}
D^{\rm mul}_n(\lambda=1)=(2m-1)(2m+3)^{n-1}+1\,.
\end{equation}

Since each resultant eigenvalue of $P_n$ inherits the multiplicity of its father, the degeneracy of every first-generation offspring of eigenvalue $1$ is $(2m-1)(2m+3)^{n-2}+1$, the degeneracy of every second-generation offspring of eigenvalue $1$ is $(2m-1)(2m+3)^{n-3}+1$, and so on. Then, the total number of eigenvalue $1$ and its descendants in $P_n$ ($n\geq 1$) is given by
\begin{eqnarray}\label{N6}
N_n(\lambda^{\rm seed}_1)&=& \sum_{i=1}^{n}
[(2m-1)(2m+3)^{i-1}+1] \times 4^{n-i}\nonumber \\
&=&(2 m+3)^n-\frac{2^{2 n+1}+1}{3} .
\end{eqnarray}

Let $N_n(\lambda^{\rm seed}_2)$ denote the number of eigenvalue $\frac{1}{m+2}$ and its descendants of $P_n$, and $N_n(\lambda^{\rm seed}_3)$ the number of eigenvalue $\frac{2m+3}{m+2}$ and its descendants of $P_n$. Then, it is obvious that
\begin{equation}\label{N7}
N_n(\lambda^{\rm seed}_2)=N_n(\lambda^{\rm seed}_3) = \sum_{i=1}^n 4^{n-i}=\frac{4^n-1}{3}\,.
\end{equation}
In addition to eigenvalues $1$, $\frac{1}{m+2}$, $\frac{2m+3}{m+2}$ and their offsprings,
$P_n$ ($n\geq 1$) has two more eigenvalues $0$ and $2$. Summing up the number of eigenvalues, we obtain
\begin{equation}\label{N8}
N_n(\lambda^{\rm seed}_1)+N_n(\lambda^{\rm seed}_2)+N_n(\lambda^{\rm seed}_3)+2
=N_n\,,
\end{equation}
which implies that we have found all eigenvalues of $P_n$.

\subsection{Applications of normalized Laplacian eigenvalues} \label{app}

We next use the above obtained eigenvalues to determine some relevant quantities for the family of
fractal trees $F_n$, including the eigentime identity
of random walks and the number of spanning trees. Since for any tree, the number of its spanning trees is 1, so our aim for computing spanning trees of $F_n$ is to verify that our computation of eigenvalues of the normalized Laplacian matrix is right.

\subsection{Eigentime identity}\label{appA}

Since the normalized Laplacian matrix of a network is derived from its transition matrix that describes isotropic discrete-time random walks on the network~\cite{MeKl04,BuCa05}, many important quantities about unbiased random walks can be extracted from the eigenvalues of normalized Laplacian matrix. For instance, the sum of reciprocals of each nonzero eigenvalue of normalized Laplacian matrix for a network determines the eigentime identity~\cite{AlFi99,LeLo02} of random walks on the network, which is a global characteristic of the network, and reflects the architecture of the whole network.

Let $H_{ij}(n)$ be the mean-first
passage time from node $i$ to node $j$ in $F_{n}$,
which is the expected time for a particle starting off from node $i$ to arrive at node $j$ for the first time~\cite{Re01,NoRi04,CoBeTeVoKl07}. The stationary distribution for random walks on $F_{n}$~\cite{Lo96,AlFi99} is $\pi=(\pi_1, \pi_2,\ldots, \pi_N)^\top$, where $\pi_i=d_i(n)/(2E_n)$, obeying relations  $\sum_{i=1}^{N}\pi_i=1$ and $\pi^{\top}M_n=\pi^{\top}$.
Let $H_n$ represent the eigentime identity for random walks on $F_{n}$, which is defined as the expected time for a walker going from a node $i$ to another node $j$, chosen randomly from all nodes accordingly to the stationary distribution. That is,
\begin{equation}\label{eig01a}
H_n=\sum_{j=1}^{N_{n}}\pi_j\,H_{ij}(n)\,.
\end{equation}

Notice that $H_n$ quantifies the expected time taken by a particle starting from node $i$ to reach a node (target) $j$ randomly chosen according the stationary distribution. Since  $H_n$ is independent of the starting node~\cite{Lo96,AlFi99}, it can be rewritten as
\begin{equation}\label{eig01b}
H_n=\sum_{i=1}^{N_{n}}\pi_i\sum_{j=1}^{N_{n}}\pi_j\,H_{ij}(n)=\sum_{j=1}^{N_{n}}\pi_j\sum_{i=1}^{N_{n}}\pi_i\,H_{ij}(n)\,.
\end{equation}
The rightmost expression in Eq.~(\ref{eig01b}) indicates that the eigentime identity $H_n$ is actually the average trapping time~\cite{Ko00} of a special trapping problem~\cite{AlFi99,TeBeVo09}, which involves a double weighted average: the former is over all the source nodes to a given trapping (target) node $j$, the latter is the average with respect to the first one taken over the stationary distribution. Because trapping is a fundamental mechanism for various other dynamical processes, $H_n$ contains much information about trapping and diverse processes taking place on complex systems, including polymer networks~\cite{WuLiZhCh12,LiZh13}.

According to previous results~\cite{AlFi99,LeLo02}, $H_n$ can be expressed in terms of the nonzero eigenvalues of $P_n$ as
\begin{equation}\label{eig01}
H_n=\sum_{i=2}^{N_{n}}\frac{1}{\lambda_{i}(n)}\,,
\end{equation}
where we have assumed that $\lambda_{1}(n)=0$. We next explicitly evaluate the sum on the right-hand side of Eq.~(\ref{eig01}).

Let $\Omega_n$ denote the set of all the $N_n-1$ nonzero eigenvalues of
$P_n$, $\Omega_n=\{\lambda_2(n), \lambda_3(n),\ldots,\lambda_{N_n}(n)\}$, where we neglect the
distinctness of the elements. We can further classify $\Omega_n$ into three nonoverlapping subsets $\Omega_n^{(1)}$, $\Omega_n^{(2)}$ and $\Omega_n^{(3)}$, satisfying $\Omega_n=\Omega_n^{(1)} \cup \Omega_n^{(2)} \cup \Omega_n^{(3)}$, where $\Omega_n^{(1)}$ only contains eigenvalue $2$, $\Omega_n^{(2)}$ consists of eigenvalue $1$ and all its offsprings, and $\Omega_n^{(3)}$ includes $\frac{1}{m+2}$, $\frac{2m+3}{m+2}$, as well as all their descendants. Then, $H_n$ can be recast as
\begin{equation}\label{eig01a}
H_n=\sum_{\lambda_{i}(n) \in \Omega_n^{(1)}}\frac{1}{\lambda_{i}(n)}          +\sum_{\lambda_{i}(n) \in \Omega_n^{(2)}}\frac{1}{\lambda_{i}(n)}+\sum_{\lambda_{i}(n) \in \Omega_n^{(3)}}\frac{1}{\lambda_{i}(n)}\,.
\end{equation}
It is obvious that $\sum_{\lambda_{i}(n) \in \Omega_n^{(1)}}\frac{1}{\lambda_{i}(n)}=\frac{1}{2}$. Hence, we only need to compute the two sum terms $\sum_{\lambda_{i}(n) \in \Omega_n^{(2)}}\frac{1}{\lambda_{i}(n)}$ and $\sum_{\lambda_{i}(n) \in \Omega_n^{(3)}}\frac{1}{\lambda_{i}(n)}$.

Equation~(\ref{T16}) shows that each eigenvalue $\lambda_i(n-1)$ in $\Omega_{n-1}^{(2)}$ (or $\Omega_{n-1}^{(3)}$) derives four eigenvalues, $\lambda_{i,1}(n)$, $\lambda_{i,2}(n)$, $\lambda_{i,3}(n)$ and $\lambda_{i,4}(n)$, which belong to $\Omega_{n}^{(2)}$ (or $\Omega_{n}^{(3)}$). According to Vieta's formulas, we have
\begin{equation}\label{eig04}
\frac{1}{\lambda_{i,1}(n)} + \frac{1}{\lambda_{i,2}(n)} + \frac{1}{\lambda_{i,3}(n)} + \frac{1}{\lambda_{i,4}(n)} = 1+\frac{6m+9}{\lambda_i(n-1)}\,
\end{equation}
and
\begin{equation}\label{eig04a}
\lambda_{i,1}(n)\times\lambda_{i,2}(n)\times\lambda_{i,3}(n)\times\lambda_{i,4}(n)=\frac{\lambda_i(n-1)}{(m+2)^2}\,.
\end{equation}
Therefore, for $n\geq 2$, we obtain
\begin{eqnarray}\label{eig05}
&\quad&\sum_{\lambda_{i}(n) \in \Omega_n^{(2)}}\frac{1}{\lambda_{i}(n)}\nonumber\\
&=&(6m+9)\sum_{\lambda_{i}(n-1) \in \Omega_{n-1}^{(2)}} \frac{1}{\lambda_{i}(n-1)}\nonumber \\
&\quad&+N_{n-1}(\lambda_1^{\rm seed})+ (2m-1)(2m+3)^{n-1}+1
\end{eqnarray}
and
\begin{eqnarray}\label{eig05b}
&\quad&\sum_{\lambda_{i}(n) \in \Omega_n^{(3)}}\frac{1}{\lambda_{i}(n)}\nonumber\\
&=&N_{n-1}(\lambda_2^{\rm seed})+N_{n-1}(\lambda_3^{\rm seed})+(m+2)+\frac{m+2}{2m+3}\nonumber\\
&\quad&+(6m+9)\sum_{\lambda_{i}(n-1) \in \Omega_{n-1}^{(3)}} \frac{1}{\lambda_{i}(n-1)}\,.
\end{eqnarray}
Then, we have the following recursive relation for $H_n$:
\begin{eqnarray}\label{eig06}
&\quad& H_n-\frac{1}{2}\nonumber\\
&=&(6m+9)\left(H_{n-1}-\frac{1}{2}\right)+N_{n-1}(\lambda_1^{\rm seed})+N_{n-1}(\lambda_2^{\rm seed})\nonumber\\
&\quad&+N_{n-1}(\lambda_3^{\rm seed})+(2m-1)(2m+3)^{n-1}+1\nonumber\\
&\quad&+(m+2)+\frac{m+2}{2m+3}\,.
\end{eqnarray}
Using Eq.~(\ref{N8}), Eq.~(\ref{eig06}) is reduced to
\begin{eqnarray}\label{eig06c}
H_n=(6m+9)H_{n-1}+\frac{2 m (2 m+3)^n-4 m^2-9 m-4}{2 m+3}\,.\nonumber\\
\end{eqnarray}

Considering $H_1=3m+\frac{5}{2}+\frac{m+2}{2m+3}$, Eq.~(\ref{eig06c}) is solved to yield
\begin{eqnarray}\label{eig07}
H_n&=&\frac{8 (m+1)^2 (6 m+9)^n-6 m^2 (2 m+3)^n}{12 m^2+34 m+24}\nonumber\\
&&-\frac{8 m (2 m+3)^n-4 m^2-9 m-4}{12 m^2+34 m+24}\,,
\end{eqnarray}
which can be expressed as a function of network size $N_{n}$ as
\begin{eqnarray}\label{eig08}
H_n&=&\frac{4 (m+1)^2}{6 m^2+17 m+12} (N_n-1)^{1+\ln 3/\ln (2 m+3)}\nonumber\\
&&-\frac{6 m^2+8 m }{12 m^2+34 m+24}N_n\nonumber\\
&&+\frac{10 m^2+17 m+4}{12 m^2+34 m+24}\,.
\end{eqnarray}
Thus, for very large networks, i.e., $N_n\rightarrow \infty$,
\begin{eqnarray}\label{eig09}
 H_{n}\sim (N_{n})^{1+\ln3/\ln(2m+3)}=(N_{n})^{1+1/f_B},
\end{eqnarray}
which grows superlinearly with the network size $N_n$, consistent with the general result in~\cite{TeBeVo09}.

\subsection{Spanning trees}

In addition to eigentime identity, the eigenvalues of normalized Laplacian matrix of a connected network also determine the number of its spanning trees. From the well established results~\cite{Ch97,ChZh07}, the number of spanning trees $N_{\rm st}(F_{n})$ for $F_n$  is
\begin{equation}\label{ST01}
N_{\rm st}(F_{n})=\frac{\displaystyle \prod_{i=1}^{N_{n}}
d_i(n)\prod_{i=2}^{N_{n}}\lambda_i(n)}{\displaystyle \sum_{i=1}^{N_{n}}d_i(n)}\,.
\end{equation}
Let $\Phi_n$, $\Psi_n$ and $\Lambda_n$ represent $\sum_{i=1}^{N_{n}}d_i(n)$, $\prod_{i=1}^{N_{n}} d_i(n)$ and $\prod_{i=2}^{N_{n}}\lambda_i(n)$, respectively. Then, the following two recursive relations hold:
\begin{equation}\label{ST02}
N_{\rm st}(F_{n})=\frac{\Psi_n \times \Lambda_n}{\Phi_n}\,
\end{equation}
and
\begin{equation}\label{ST03}
N_{\rm st}(F_{n-1})=\frac{\Psi_{n-1} \times \Lambda_{n-1}}{\Phi_{n-1}}\,,
\end{equation}
from which we can derive the relation between $N_{\rm st}(F_{n})$ and $N_{\rm st}(F_{n-1})$.

Obviously,
\begin{equation}\label{ST04}
\Phi_n=\sum_{i=1}^{N_{n}}d_i(n)=2E_n=2(2m+3)^n\,,
\end{equation}
leading to
\begin{equation}\label{ST05}
\Phi_n=(2m+3)\Phi_{n-1}\,.
\end{equation}
On the other hand, by the construction and structural properties of the fractal trees, we have
\begin{equation}\label{ST06}
\Psi_n=(m+2)^{2E_{n-1}}\Psi_{n-1}\,.
\end{equation}
Finally,
\begin{equation}\label{ST07}
\Lambda_n=2\prod_{\lambda_{i}(n) \in \Omega_n^{(2)}}\lambda_{i}(n)\,\prod_{\lambda_{i}(n) \in \Omega_n^{(3)}}\lambda_{i}(n)\,,
\end{equation}
where the two product terms on the right-hand side satisfy
\begin{eqnarray}\label{ST08}
&\quad&\prod_{\lambda_{i}(n) \in \Omega_n^{(2)}}\lambda_{i}(n)\nonumber\\
&=&\left(\frac{1}{m+2}\right)^{2 N_{n-1}(\lambda_1^{\rm seed})} \prod_{\lambda_{i}(n-1) \in \Omega_{n-1}^{(2)}} \lambda_{i}(n-1)
\end{eqnarray}
and
\begin{eqnarray}\label{ST09}
\prod_{\lambda_{i}(n) \in \Omega_n^{(3)}}\lambda_{i}(n)&=&\left(\frac{1}{m+2}\right)^{2 \left[N_{n-1}(\lambda_2^{\rm seed})+N_{n-1}(\lambda_3^{\rm seed})\right]}\times \nonumber\\
&& \frac{2m+3}{(m+2)^2}  \prod_{\lambda_{i}(n-1) \in \Omega_{n-1}^{(3)}}\lambda_{i}(n-1) \,.
\end{eqnarray}
Thus,
\begin{eqnarray}\label{ST10}
\Lambda_n&=&\left(\frac{1}{m+2}\right)^{2 \left[N_{n-1}(\lambda_2^{\rm seed})+N_{n-1}(\lambda_2^{\rm seed})+N_{n-1}(\lambda_3^{\rm seed})\right]}\nonumber\\
&\quad&\times \frac{2m+3}{(m+2)^2} \times \Lambda_{n-1}\nonumber\\
&=&(2m+3)\left(\frac{1}{m+2}\right)^{2E_{n-1}}\times \Lambda_{n-1}\,.
\end{eqnarray}
Equations~(\ref{ST05}),~(\ref{ST06}), and~(\ref{ST10}) indicate that
\begin{eqnarray}\label{eig06d}
N_{\rm st}(F_{n})=N_{\rm st}(F_{n-1})=1\,,
\end{eqnarray}
which implies that our computation for eigenvalues of the normalized Laplaican matrix $P_n$ is correct.

\section{Normalized Laplacian spectrum for Cayley trees}

This section is devoted to the eigenvalue problem and their applications of normalized Laplacian matrices for Cayley trees. Here, we focus on a particular Cayley tree corresponding to $b=3$, since for other values of $b$ the computation and result is similar but the formulas are very lengthy.

\subsection{Characteristic polynomial and eigenvalues}

Since the above method for computing the spectrum of the normalized Laplacian matrix for the fractal trees is not applicable to the Cayley trees, we use the elementary matrix operations to reduce the related matrix to lower triangle matrix.

For the convenience description, we use the same notations as those of the fractal trees. Let $A_n$ and $D_n$ be the adjacency matrix and diagonal degree matrix of $C_{3,n}$. Then, its normalized Laplacian matrix is $P_n=I_n-D_n^{-\frac{1}{2}} A_n D_n^{-\frac{1}{2}}$, which has the same set of eigenvalues as matrix $I_n-D_n^{-1}A_n$. Below we concentrate on matrix $I_n-D_n^{-1}A_n$ and still denote it as $P_n$ in the case without confusion.

Note that the numbers of nodes and edges in $C_{3,n}$ are $N_n=3\times 2^n-2$ and $E_n=3 \times 2^n-3$, respectively.
The $N_n=3\times 2^n-2$ nodes can be divided into $n+1$ levels: the $0$th level contains only one node (i.e., the central node) labeled by $1$; the $i$th ($1\leq i \leq n$) level has $N_i(n)=N_i-N_{i-1}$ nodes, which are labeled sequentially by $N_{i-1}+1, N_{i-1}+2,\cdots, N_i$.
Here $N_i$ is defined by Eq.~(\ref{cay1a}).

We now address the eigenvalue problem of $P_n$. By definition, all eigenvalues of $P_n$ are actually the roots of characteristic equation $\det (P_n-\lambda I_n)=0$. Let $X_n=P_n-\lambda I_n$. Then $\det X_n$ is a determinant of order $3\times 2^n-2$. Next we apply the row operations of determinants to transform $\det X_n$ into a lower triangle determinant.

Let $R_k$ represent the $k$th row of $X_n$ and its variants after being performed row operations. In order to have a lower triangle matrix, the row operations are performed as follows. First, we keep rows $R_k$ ($N_{n-1}+1 \leq k \leq N_n $) unchanged, and define their diagonal entries as ${f_1}(\lambda)= 1-\lambda$ as in the original matrix. Then for each $1 \leq i \leq n$, we repeat the following two operations: (i) For each $k$ ($N_{n-i-1}+1 \leq k \leq N_{n-i}$), we multiply $R_k$ by ${f_i}(\lambda)$; (ii) For each $k$ ($N_{n-i-1}+1 \leq k \leq N_{n-i}$), we add the sum of $R_{2k+1}$ and $R_{2k+2}$ times $\frac{1}{3}$ to $R_k$. Note that we have assumed $N_{-1}=0$. After performing the two operations, for $i < n$, the diagonal entry of $R_k$ ($N_{n-i-1}+1 \leq k \leq N_{n-i}$) becomes $f_{i+1}(\lambda)$, while the other entries of $R_k$ on the right-hand side of the diagonal entry $f_{i+1}(\lambda)$ are zeros. Finally, we add $R_2$ times $\frac{1}{3}$ to $R_1$.

The above row operations reduce matrix $X_n$ to lower triangle matrix $Y_n$, the diagonal elements of which are as follows: ${f_{n+1}(\lambda)}-\frac{1}{9}{f_{n-1}(\lambda)}$ for the first row, ${f_i}(\lambda)$ ($1 \leq i \leq n$) for those rows starting from $N_{n-i}+1$ to $N_{n-i+1}$. It follows above row operations that the functions ${f_i}(\lambda)$ obey the following recursive relations:
\begin{equation}\label{DetPn1}
{f_i}(\lambda)=\begin{cases}
1-\lambda, &i=1,\\
{\lambda ^2}-2\lambda+\frac{1}{3}, &i=2,\\
(1-\lambda){f_{i-1}}(\lambda)-\frac{2}{9}{f_{i-2}}(\lambda), &3\leq i\leq n+1.
\end{cases}
\end{equation}

According to the properties of determinants, we have $\det X_n=\det Y_n/D(\lambda)$, where $D(\lambda)$ is the overall factor. From the above procedure, we can obtain
\begin{equation}
D(\lambda) ={f_n}(\lambda )\prod\limits_{i=1}^{n-1} {\left[{f_i}(\lambda )\right]^{3\times{2^{n-i-1}}}} \,.
\end{equation}
Thus, we have
\begin{eqnarray}\label{DetPn2}
&&\det X_n\nonumber\\
&=&\frac{[{f_n}(\lambda)]^3[{f_{n + 1}}(\lambda ) - \frac{1}{9}{f_{n - 1}}(\lambda )]\prod\limits_{i=1}^{n-1} [{f_i}(\lambda )]^{3\times{2^{n-i}}}}{D_n(\lambda )} \nonumber\\
  &=& [{f_n}(\lambda)]^2[{f_{n + 1}}(\lambda ) - \frac{1}{9}{f_{n - 1}}(\lambda )]\prod\limits_{i=1}^{n-1} [{f_i}(\lambda )]^{3\times{2^{n-i-1}}} \,.\nonumber\\
\end{eqnarray}

Since the eigenvalues of $P_n$ are the roots of $\det X_n=0$ and the right-hand side of Eq.~(\ref{DetPn2}) is factorized, the problem of computing eigenvalues of $P_n$ becomes to find the roots of functions $f_i(\lambda)$ ($1\leq i\leq n$) and $f_{n + 1}(\lambda ) - \frac{1}{9}f_{n - 1}$. From Eq.~(\ref{DetPn1}), it is obvious that $f_i(\lambda)$ is a polynomial of $\lambda$ with degree $i$. Every $f_i(\lambda)$ provides $i$ different roots, each of which is an eigenvalue of $P_n$ having a multiplicity $3\times{2^{n-i-1}}$, with the exception of $f_n(\lambda)$, whose roots are  2-fold degenerate. In addition, $f_{n + 1}(\lambda )-\frac{1}{9}f_{n - 1}$ generates $n+1$ eigenvalues. Thus the total number of eigenvalues is
\begin{equation}
2n+(n+1)+\sum_{i=1}^{n-1}(3i\times 2^{n-i-1})=3\times 2^n-2=N_n\,,
\end{equation}
implying that all eigenvalues of $P_n$ are produced by $f_i(\lambda)=0$ ($i=1,2,\ldots,n$) and $f_{n + 1}(\lambda )-\frac{1}{9}f_{n - 1}=0$.

\subsection{Applications of normalized Laplacian eigenvalues} \label{app2}

Although we failed to obtain the explicit recursive expressions for the eigenvalues of $P_n$, below we will show that the above obtained characteristic polynomial for $P_n$ contains enough information of the eigenvalues such that we can use it to determine the eigentime identity and the number of spanning trees in Cayley trees.

\subsection{Eigentime identity}

For each $f_i(\lambda)$ ($1\leq i\leq n$), it is an $i-$degree polynomial, and  we can rewrite it as:
\begin{equation}\label{DetPn3}
f_i(\lambda)=\alpha_i + {\beta _i}\lambda + {\gamma _i}{\lambda^2} + \cdots\,.
\end{equation}
Thus, equation $f_i(\lambda)=0$ has $i$ nonzero roots, labeled by $\lambda_1^{(i)}$, $\lambda_2^{(i)}$, $\cdots$, $\lambda_i^{(i)}$. According to Vieta's formulas, we have
\begin{equation}\label{DetPn4}
\sum\limits_{j = 1}^{i} \frac{1}{\lambda_j^{(i)}}=- \frac{{{\beta_i}}}{{{\alpha_i}}}\,,
\end{equation}
and
\begin{equation}\label{DetPn5}
\prod\limits_{j = 1}^{i} \lambda_j^{(i)}=\alpha_i\,,
\end{equation}
where we have used the fact that the coefficient of the term $\lambda^i$ in $f_i(\lambda)$ equals $(-1)^i$.

From the recursion relation in Eq.~(\ref{DetPn1}), we obtain
\begin{equation}\label{DetPn6}
\alpha_{i + 1}={\alpha _i} - \frac{2}{9}{\alpha _{i - 1}}\,,
\end{equation}
\begin{equation}\label{DetPn7}
\beta_{i + 1}= {\beta _i} - {\alpha _i} - \frac{2}{9}{\beta _{i - 1}}\,,
\end{equation}
and
\begin{equation}\label{DetPn8}
\gamma_{i + 1} = {\gamma _i} - {\beta _i} - \frac{2}{9}{\gamma _{i - 1}}\,.
\end{equation}
Considering the initial conditions ${\alpha _1} = 1$, ${\beta _1} =  - 1$, ${\gamma _1} = 0$, ${\alpha_2} = \frac{1}{3}$, ${\beta_2} =-2$ and ${\gamma_2}=1$, Eqs.~(\ref{DetPn6}-\ref{DetPn8}) are solved to yield
\begin{equation}
{\alpha _i}= {3^{-(i-1)}} \,,
\end{equation}
\begin{equation}
\beta_i =  - 3^{-(i-1)}\left(2^{i+2} - 3i- 4\right)\,,
\end{equation}
and
\begin{equation}
\gamma_i = \frac{1}{2}3^{-(i-2)}\left(2^{i+3} i- 9\times{2^{i+2}} + 3{i^2}+ 17i+36\right)\,.
\end{equation}
Therefore, for any $1\leq i\leq n$,
\begin{equation}
\sum \limits_{j = 1}^{i} {\frac{1}{{\lambda_j^{(i)}}}}  =  - \frac{{{\beta _i}}}{{{\alpha _i}}} = 2^{i+2}- 3i- 4\,,
\end{equation}
and
\begin{equation}
\prod\limits_{j = 1}^{i} \lambda_j^{(i)}  = 3^{-(i-1)} \,,
\end{equation}

On the other hand, we can see that the only zero eigenvalue of $P_n$ is generated by $f_{n + 1}(\lambda) - \frac{1}{9}f_{n - 1}(\lambda)=0$, since
\begin{eqnarray}
&&f_{n + 1}(\lambda) - \frac{1}{9}f_{n - 1}(\lambda) \nonumber\\
&=& \lambda\left[ \left(\beta _{n + 1} -\frac{1}{9}\beta_{n - 1}\right) +\left(\gamma_{n + 1}- \frac{1}{9}\gamma_{n - 1}\right)\lambda + \cdots\right]\,,\nonumber\\
\end{eqnarray}
which implies that the $n$ nonzero roots, $\lambda_1^{(n+1)}$, $\lambda_2^{(n+1)}$, $\cdots$, $\lambda_n^{(n+1)}$, satisfy
\begin{equation}
\sum\limits_{j = 1}^{n} {\frac{1}{{\lambda_j^{(n + 1)}}} =  - \frac{{{\gamma _{n + 1}} - \frac{1}{9}{\gamma _{n - 1}}}}{{{\beta _{n + 1}} - \frac{1}{9}{\beta _{n - 1}}}}}\,.
\end{equation}
and
\begin{equation}
\prod\limits_{j =1}^{n} \lambda_j^{(n + 1)} = \frac{1}{9}{\beta _{n - 1}}-{\beta _{n + 1}}\,.
\end{equation}

Let $\lambda_j(n)$ ($j=1,2,\ldots, N_n$) denote the $N_n$ eigenvalues of the normalized Laplacian matrix $P_n$ of the Cayley tree $C_{3,n}$, where we assume that $\lambda_1(n)=0$. Then, the eigentime identity for random walks on $C_{3,n}$ is given by
\begin{eqnarray}
H_n&=&\sum\limits_{j = 2}^{N_n} \frac{1}{\lambda_j(n)}  \nonumber\\
 &=& -\left[\sum\limits_{i = 1}^{n- 1} {3\times{2^{n - i - 1}}\times \frac{{{\beta _i}}}{{{\alpha _i}}}} \right]-2 \times  \frac{{{\beta _n}}}{{{\alpha _n}}}\nonumber\\
 &&-\frac{\gamma _{n + 1} - \frac{1}{9}\gamma _{n - 1}}{\beta _{n + 1} - \frac{1}{9}\beta _{n - 1}}\nonumber\\
  &=& \frac{{ 3n\times{4^{n+1}} - 13\times{2^{2n+1}}+ 35\times{2^n}- 9  }}{{2({2^n}- 1)}}\,,
\end{eqnarray}
which can be expressed in terms of network size $N_n$ as
\begin{eqnarray}
H_n&=&\frac{2(N_n)^2 \ln (N_n+2)}{(N_n-1)\ln 2}-\frac{(N_n)^2 (13 \ln 2 + 6\ln 3)}{3(N_n-1)\ln 2}\nonumber\\
&&+\frac{8 N_n \ln (N_n+2)}{(N_n-1) \ln 2}+\frac{N_n (\ln 2-48 \ln 3)}{6 (N_n-1) \ln 2}\nonumber\\
&&+\frac{8 \ln (N_n+2)}{(N_n-1) \ln 2}
 +\frac{25 \ln 2-48 \ln 3}{6 (N_n-1) \ln 2}\,.
\end{eqnarray}
For very large networks (i.e., $N_n\to \infty$), the leading term of $H_n$ obey
\begin{equation}
H_n\sim N_n\ln N_n\,,
\end{equation}
which is in sharp contrast to its counterpart of the fractal trees $F_n$ obtained in Sec.~\ref{appA}. Note that for the case $b>3$, the dominating term of $H_n$ follows the same scaling as that corresponding to the case $b=3$.

\subsection{Spanning trees}

From above results, we can also obtain the product of nonzero eigenvalues for $P_n$: \begin{eqnarray}
&&\prod\limits_{j = 2}^{N_n} \lambda _j(n) \nonumber\\
 &=& \left[\prod\limits_{i = 1}^{n- 1} (\alpha_i)^{3\times{2^{n - i - 1}}} \right] \times (\alpha_n)^2\times \left(\frac{1}{9}{\beta _{n - 1}}-{\beta _{n + 1}}\right)\nonumber\\
  &=& 2 \left(2^n-1\right)3^{-\frac{3}{2} \left(2^n-2\right)}\,,
\end{eqnarray}
In addition, we can derive
\begin{equation}
\prod_{i=1}^{N_n} d_i(n)=3^{N_{n-1}}=3^{3\times 2^{n-1}-2}\,,
\end{equation}
and
\begin{equation}
\sum_{i=1}^{N_n} d_i(n)=3N_{n-1}+(N_n-N_{n-1})=6\times 2^n-6\,.
\end{equation}
Then, the number of spanning trees in $C_{3,n}$ is
\begin{equation}\label{StCayley}
N_{\rm st}(C_{3,n})=\frac{\displaystyle \prod_{i=1}^{N_{n}}
d_i(n)\prod_{i=2}^{N_{n}}\lambda_i(n)}{\displaystyle \sum_{i=1}^{N_{n}}d_i(n)}=1\,,
\end{equation}
which means that our computation about the characteristic polynomial and eigenvalues for related matrix of $C_{3,n}$ is valid.

\section{Conclusions}

The eigenvalue spectrum of the normalized Laplacian matrix of a network is relevant in the topological aspects and random-walk dynamics that is closely related to a large variety of other dynamical processes of the network. In this paper, we have studied the eigenvalue problem of the normalized Laplacian matrices of a class of fractal trees and Cayley trees, with the latter being a typical model of dendrimers. Both networks under study are constructed in an iterative way, which allows to analytically treat the eigenvalues and their degeneracies.

For the fractal trees, by applying the decimation technique, we have provided an exact recursive relation governing the eigenvalues of the networks at two consecutive generations. Then, we have also derived the multiplicity of each eigenvalue. For the Cayley trees, through the elementary matrix operations we have reduced the related matrix to lower triangle matrix and reduced the eigenvalue problem to computing the roots of some polynomials of very small degrees, with the polynomials being defined recursively.

On the basis of these obtained recursion relations of eigenvalues for fractal trees and of related polynomials for Cayley trees, we have further evaluated the eigentime identity for random walks on the two networks, the leading scalings of which display disparate behaviors. In addition, we have verified our computation and results for the eigenvalues of the normalized Laplacian matrices for the two different types of networks by enumerating the number of spanning trees in them.

The fractal trees being studied have a similar topology to the Vicsek fractals modeling regular hyperbranched polymers, both fractals are thus suggested to exhibit similar dynamical behaviors, e.g., superlinear growth of eigentime identity. Since eigentime identity is an important quantity rooted in the inherent network topology, the distinct scalings of eigentime identity enable to distinguish two classic macromolecules---dendrimers and hyperbrached polymers. On the other hand,  although we have limited our analysis to unweighted treelike networks, the approaches and procedures applied here could be extended to the case of weighted~\cite{BaKl98,BaKl98JCP} and directed networks, even those with loops.

\begin{acknowledgments}
We would like to thank Hongxiao Liu and Zhengyi Hu for assistance. This work was supported by the National Natural Science Foundation of China under Grant Nos. 61074119 and 11275049. Bin Wu also acknowledged the support of the Graduate Student Innovation Foundation of Fudan University.
\end{acknowledgments}

\appendix*

\section{Proof of equation~(\ref{T7})\label{AppA}}

The $N_{n+1}-N_{n}$ nodes in set $\beta$ can be divided into $E_{n}$ sections. All nodes in each section associates an edge in $F_{n}$ that generates these nodes. Concretely, each section contains $2(m+1)$ nodes, among which two nodes have a degree of $m+2$, the other $2m$ nodes have a single degree. The two nodes with degree $m+2$ are connected to each other by an edge; moreover, each multiple-degree node is linked to $m$ single-degree nodes in the same section. Notice that there is no edge between any pair of nodes belonging to different sections. Hence, $P_{\beta,\beta}$ is a block diagonal matrix including $E_{n}$ blocks, with each block having the form \begin{equation} \label{Note1}
\left[\begin{array}{cccccccc}
    1 & -\frac{1}{\sqrt{d}} & \dots & -\frac{1}{\sqrt{d}} & -\frac{1}{d} & 0 & \dots & 0 \\
    -\frac{1}{\sqrt{d}} & 1 & \dots & 0 &  0 & 0 & \dots &  0\\
    \vdots & \vdots & \ddots & \vdots & \vdots & \vdots & \ddots & \vdots  \\
    -\frac{1}{\sqrt{d}} & 0 & \dots & 1 & 0 & 0 & \dots & 0 \\
    -\frac{1}{d} & 0 & \dots & 0 & 1 & -\frac{1}{\sqrt{d}} & \dots  & -\frac{1}{\sqrt{d}} \\
    0 & 0  & \dots & 0 & -\frac{1}{\sqrt{d}} &1  & \dots & 0  \\
    \vdots & \vdots & \ddots & \vdots & \vdots & \vdots & \ddots &\vdots  \\
    0 & 0 & \dots & 0  & -\frac{1}{\sqrt{d}} & 0 & \dots & 1
\end{array}\right],
\end{equation}
where $d=m+2$ represents the degree of the two non-single-degree nodes, denoted by $k$ and $l$, respectively.

According to above arguments, we can conclude that $T_n =[\lambda_i(n + 1)-P_{\beta,\beta}]^{-1}$ is also block diagonal, with each of $E_{n}$ blocks being identical. It is not difficult to derive that their entries satisfy relations $T_n(k,k)=T_{n}(l,l)$ and $T_{n}(l,k)=T_{n}(k,l)$, the exact expressions of which are
\begin{widetext}
\begin{eqnarray}\label{Note3a}
T_n(k,k)=\frac{(m+2)^2 [\lambda_i(n+1)]^3-3 (m+2)^2 [\lambda_i(n+1)] ^2+\left(2 m^2+10 m+12\right) \lambda_i(n+1) -2 m-4}{(m+2)^2 [\lambda_i(n+1)]^4-4 (m+2)^2 [\lambda_i(n+1)]^3+\left(4 m^2+20 m+23\right) [\lambda_i(n+1)]^2-(8 m+14)\lambda_i(n+1) +3}\,,
\end{eqnarray}
\begin{eqnarray}\label{Note3b}
T_{n}(l,k)=\frac{-(m+2) [\lambda_i(n+1)] ^2+(2m+4) \lambda_i(n+1) -m-2}{(m+2)^2 [\lambda_i(n+1)]^4-4 (m+2)^2 [\lambda_i(n+1)]^3+\left(4 m^2+20 m+23\right) [\lambda_i(n+1)]^2-(8 m+14) \lambda_i(n+1) +3}\,,
\end{eqnarray}
\end{widetext}
which are useful to derive Eq.~(\ref{T16}) in the main text.

We proceed to determine the entries $Q_n(i,j)$ of $Q_n=I_n+P_{\alpha,\beta}[\lambda_i(n + 1)-P_{\beta,\beta}]^{-1} P_{\beta,\alpha}$. By definition,
\begin{eqnarray} \label{AApp1}
&&Q_n(i,j) =\delta_{i,j}+\nonumber \\
&&\frac{1}{\sqrt{d_i(n+1)d_j(n+1)}} \sum_{\substack{u \in \beta, v \in \beta \\ i\thicksim u, j\thicksim v }} \frac{T_{n}(u,v)}{\sqrt{d_u(n+1)d_v(n+1)}},\nonumber \\
\end{eqnarray}
where $i\thicksim u$ means that nodes $i$ and $u$ are adjacent in $F_{n+1}$, and $\delta_{i,j}$ is defined as: $\delta_{i,j}=1$ if $i=j$, $\delta_{i,j}=0$ otherwise. Note that for nodes in set $\beta$, only those with multiple degrees have a link connected to an old node in set $\alpha$. Then, $d_u(n+1)=d_v(n+1)=m+2$ and Eq.~(\ref{AApp1}) becomes
\begin{small}
\begin{equation} \label{AApp2}
Q_n(i,j)=\delta_{i,j}+\frac{1}{(m + 2)\sqrt{d_i(n+1)d_j(n+1)}} \sum_{\substack{u \in \beta, v \in \beta \\ i\thicksim u, j\thicksim v }} T_{n}(u,v).
\end{equation}
\end{small}

Equation~(\ref{AApp2}) can be simplified by distinguishing two cases: $i=j$ and $i \neq j$.
For the case $i=j$, we have
\begin{equation} \label{AApp3}
Q_n(i,i) =1+\frac{d_i(n+1)}{(m+2)d_i(n+1)} T_n(u,u)=1+\frac{T_n(l,l)}{m+2}\,,
\end{equation}
For the other case $i \neq j$,
\begin{equation} \label{AApp5}
Q_n(i,j)= \frac{T_n(u,v)}{m+2} \frac{1}{\sqrt{d_i(n+1)d_j(n+1)}}\,.
\end{equation}
Moreover, if nodes $i$ and $j$ are not adjacent in $F_n$, the non-diagonal elements $Q_n(i,j)$ of $Q_n$ are zeros, because in this case $u$ and $v$ are in different sections, leading to $T_n(u,v)=0$. Otherwise, if nodes $i$ and $j$ are directly connected by an edge in $F_n$, then $T_n(u,v)=T_n(l,k)$ with the latter given by Eq.~(\ref{Note3b}). Notice that for any old node $i$ in $\alpha$, we have $d_i(n)=d_i(n+1)$. Thus, Eq.~(\ref{AApp5}) can be recast as
\begin{equation} \label{AApp6}
Q_n(i,j)= \frac{T_n(l,k)}{m+2} \frac{A_n(i,j)}{\sqrt{d_i(n)d_j(n)}}=-\frac{T_n(l,k)}{m+2}P_n(i,j)\,.
\end{equation}
Combining the results provided by Eq.~(\ref{AApp3}) and~(\ref{AApp6}), we can obtain
\begin{eqnarray} \label{App6}
Q_n=-\frac{T_n(l,k)}{m+2} P_n+\left(1+\frac{T_n(l,l)+T_n(l,k)}{m+2}\right)I_n\,.
\end{eqnarray}
This completes the proof of Eq.~(\ref{T7}).

\nocite{*}

\begin{references}


\bibitem{GuBl05}
A. A. Gurtovenko and A. Blumen,
Adv. Polym. Sci. {\bf 182}, 171 (2005).

\bibitem{BoLaMoChHw06}
S. Boccaletti, V. Latora, Y. Moreno, M. Chavez and D.-U. Hwanga,
Phys. Rep. {\bf 424}, 175 (2006).

\bibitem{DoGoMe08}
S. N. Dorogovtsev, A. V. Goltsev and J.F.F. Mendes,
Rev. Mod. Phys. {\bf 80}, 1275 (2008).

\bibitem{Wu04}
F. Y. Wu, J. Phys. A: Math. Gen. {\bf 37}, 6653 (2004).

\bibitem{SoBl95}
J. U. Sommer and A. Blumen,
J. Phys. A {\bf 28}, 6669 (1995).

\bibitem{Sc98}
H. Schiessel,
Phys. Rev. E {\bf 57}, 5775 (1998).


\bibitem{BlVoJuKo05JOL}
A. Blumen, A. Volta, A. Jurjiu, and Th. Koslowski,
J. Lumin. {\bf 111}, 327 (2005).

\bibitem{BlVoJuKo05}
A. Blumen, A. Volta, A. Jurjiu, and Th. Koslowski,
Physica A {\bf 356}, 12 (2005).

\bibitem{LiuZh13}
H. X. Liu and Z. Z. Zhang,
J. Chem. Phys. {\bf 138}, 114904 (2013).

\bibitem{AhDaZa93}
Y. Aharonov, L. Davidovich, and N. Zagury,
Phys. Rev. A {\bf 48}, 1687 (1993).

\bibitem{AgBlMu08}
E. Agliari, A. Blumen, and O. M\"{u}lken,
J. Phys. A {\bf 41}, 445301 (2008).

\bibitem{MuBl11}
 O. M\"{u}lken and A. Blumen, Phys. Rep. {\bf 502}, 37 (2011).

\bibitem{DejoMe98}
A. I. M. Denneman, R. J. J. Jongschaap, and J. Mellema,
J. Eng. Math. {\bf 34}, 75 (1998).

\bibitem{CoKa92}
M. G. Cosenza and R. Kapral, Phys. Rev. A 46, 1850 (1992).

\bibitem{MaMaPe97}
U. Marini, B. Marconi, and A. Petri, J. Phys. A {\bf 30}, 1069 (1997).

\bibitem{BlJu02}
A. Blumen and A. Jurjiu,
J. Chem. Phys. {\bf 116}, 2636 (2002).

\bibitem{JaWuCo92}
C. S. Jayanthi, S. Y. Wu, and J. Cocks,
Phys. Rev. Lett. {\bf 69}, 1955 (1992).

\bibitem{JaWu94}
C. S. Jayanthi and S. Y. Wu,
Phys. Rev. B {\bf 50}, 897 (1994).

\bibitem{JuVoBe11}
A. Jurjiu, A. Volta, and T. Beu,
Phys. Rev. E {\bf 84}, 011801 (2011).

\bibitem{CaCh97}
C. Cai and Z. Y. Chen,
Macromolecules {\bf 30}, 5104 (1997).

\bibitem{GaBl07}
M. Galiceanu and A. Blumen,
J. Chem. Phys. {\bf 127}, 134904 (2007).

\bibitem{Ga10}
M. Galiceanu,
J. Phys. A {\bf 43}, 305002 (2010).

\bibitem{ZhZhQiGu08}
Z. Z. Zhang, S. G. Zhou, Y. Qi, and J. H. Guan,
Eur. Phys. J. B {\bf 63}, 507 (2008).

\bibitem{ZhQiZhLiGu09}
Z. Z. Zhang, Y. Qi, S. G. Zhou, Y. Lin, and J. H. Guan,  Phys. Rev.
E, {\bf 80}, 016104 (2009).

\bibitem{GrGrTi12}
C Grabow, S. Grosskinsky, and M. Timme,
Phys. Rev. Lett., {\bf 108}, 218701 (2012).

\bibitem{Ch97}
F. Chung, {\it Spectral Graph Theory} (American Mathematical Society, Providence, RI 1997).

\bibitem{ChZh07}
H. Y. Chen and F. J. Zhang, Discrete Appl. Math. {\bf 155}, 654
(2007).


\bibitem{Lo96}
L. Lov\'az, \emph{Random Walks on Graphs: A Survey}, in Combinatorics,
Paul Erd\"{o}s is Eighty Vol. 2, edited by D. Mikl\'os, V.
T. S\'o, and T. Sz\"{o}nyi (J\'aos Bolyai Mathematical Society,
Budapest, 1996), pp. 353-398; http://research.microsoft.com/
users/lovasz/erdos.ps.

\bibitem{AlFi99}
D. Aldous and J. Fill, Reversible Markov chains and random walks on
graphs, 1999, http://www.stat.berkeley.edu/~aldous/RWG/book.html

\bibitem{ZhShCh13}
Z. Z. Zhang, T. Shan, and G. R. Chen,
Phys. Rev. E {\bf 87}, 012112 (2013).

\bibitem{KeSn76}
J. G. Kemeny and J. L. Snell, \emph{Finite Markov Chains} (Springer,
New York, 1976).

\bibitem{LeLo02}
M. Levene and G. Loizou,
Am. Math. Mon. {\bf 109}, 741 (2002). 




\bibitem{BaKlKo97}
A. Bar-Haim, J. Klafter, and R. Kopelman, J. Am. Chem. Soc.
{\bf 119}, 6197  (1997).


\bibitem{Kn68}
R.S. Knox,
J. Theoret. Biol. {\bf 21}, 244 (1968)

\bibitem{BlZu81}
A. Blumen and G. Zumofen,
J. Chem. Phys. {\bf 75}, 892 (1981).


\bibitem{Ki58}
S. K. Kim,
J. Chem. Phys. {\bf 28}, 1057 (1958).


\bibitem{MoSh57}
E. W. Montroll and K. E. Shuler,
Adv. Chem. Phys. {\bf 1},  361 (1957).


\bibitem{We67}
G. H. Weiss,
Adv. Chem. Phys. {\bf 13}, 1 (1967).


\bibitem{ChDaHaLiPaSt04}
G. T. Chen, G. Davis, F. Hall, Z. S. Li, K. Patel, and M. Stewart,
SIAM J. Discrete Math. {\bf 18}, 353 (2004).

\bibitem{BlJuKoFe03}
A. Blumen, A. Jurjiu, Th. Koslowski, and Ch. von Ferber, Phys. Rev.
E {\bf 67}, 061103 (2003).

\bibitem{BlFeJuKo04}
A. Blumen, Ch. von Ferber, A. Jurjiu, and Th. Koslowski,
Macromolecules {\bf 37}, 638 (2004).

\bibitem{ZhWjZhZhGuWa10}
Z. Z. Zhang, B. Wu, H. J. Zhang, S. G. Zhou, J. H. Guan, and Z. G.
Wang, Phys. Rev. E {\bf 81}, 031118 (2010).

\bibitem{ChLuVu03}
F. Chung, L. Lu, and V. Vu, Proc. Natl. Acad. Sci. U.S.A. {\bf 100},
6313 (2003).

\bibitem{HwYuLeKa10}
S. Hwang, C.-K. Yun, D.-S. Lee, B. Kahng, and D. Kim,
Phys. Rev. E {\bf 82}, 056110 (2010).

\bibitem{KiKa07}
D. Kim and B. Kahng,
%
Chaos {\bf 17}, 026115 (2007).

\bibitem{WuZh12}
S. Q. Wu and Z. Z. Zhang,
J. Phys. A: Math. Theor. {\bf 45}, 345101 (2012).

\bibitem{ChCa99}
Z. Y. Chen and C. Cai,
Macromolecules {\bf  32}, 5423 (1999).

\bibitem{DoAlBeKa83}
E. Domany, S. Alexander, D. Bensimon, and L. P. Kadanoff,
Phys. Rev. B {\bf 28}, 3110 (1983).

\bibitem{Ra84}
R. Rammal, J. Phys. (Paris) {\bf 45}, 191 (1984).


\bibitem{MeKl04}
R. Metzler and J. Klafter, J. Phys. A {\bf 37}, R161 (2004).

\bibitem{BuCa05}
R Burioni and D Cassi, J. Phys. A {\bf 38}, R45 (2005).

\bibitem{Re01}
S. Redner, \emph{A Guide to First-Passage Processes} (Cambridge
University Press, Cambridge, England 2001).

\bibitem{NoRi04}
J. D. Noh and H. Rieger, Phys. Rev. Lett. {\bf 92}, 118701 (2004).

\bibitem{CoBeTeVoKl07}
S. Condamin, O. B\'enichou, V. Tejedor, R. Voituriez, and J.
Klafter,
Nature (London) {\bf 450}, 77 (2007).


\bibitem{Ko00}
J. J. Kozak,
Adv. Chem. Phys. {\bf 115}, 245 (2000).


\bibitem{TeBeVo09}
V. Tejedor, O. B\'enichou, and R. Voituriez, Phys. Rev. E {\bf 80},
065104(R) (2009).


\bibitem{WuLiZhCh12}
B. Wu, Y. Lin, Z. Z. Zhang, G. R. Chen
J. Chem. Phys. {\bf 137}, 044903 (2012).



\bibitem{LiZh13}
Y. Lin and Z. Z. Zhang,
J. Chem. Phys. {\bf 138}, 094905 (2013).

\bibitem{BaKl98}
A. Bar-Haim and J. Klafter, J. Phys. Chem. B {\bf 102}, 1662  (1998).

\bibitem{BaKl98JCP}
A. Bar-Haim and J. Klafter, J. Chem. Phys. {\bf 109}, 5187  (1998).


\end{references}

\end{document}